\documentclass[nofootinbib,twocolumn,preprintnumbers]{revtex4-1}
\pdfoutput=1

\usepackage{amsmath,amssymb,graphicx,booktabs,bm,psfrag,color}

\begin{document}

\title{\boldmath The ``forgotten'' decay $S\to Zh$ as a CP analyzer}

\preprint{MITP/16-067}
\preprint{July 4, 2016}

\author{Martin Bauer$^a$}
\author{Matthias Neubert$^{b,c}$}
\author{Andrea Thamm$^b$}

\affiliation{$^a$Institut f\"ur Theoretische Physik, Universit\"at Heidelberg, Philosophenweg 16, 69120 Heidelberg, Germany\\
${}^b$PRISMA Cluster of Excellence {\em\&} MITP, Johannes Gutenberg University, 55099 Mainz, Germany\\
${}^c$Department of Physics {\em\&} LEPP, Cornell University, Ithaca, NY 14853, U.S.A.}

\begin{abstract}
Scalar particles $S$ which are gauge singlets under the Standard Model are generic features of many models of fundamental physics, in particular as possible mediators to a hidden or dark sector. We show that the decay $S\to Zh$ provides a powerful probe of the CP nature of the scalar, because it is allowed only if $S$ has CP-odd interactions. We perform a model-independent analysis of this decay in the context of an effective Lagrangian and compute the relevant Wilson coefficients arising from integrating out heavy fermions to one-loop order. We illustrate our findings with the example of the 750\,GeV diphoton resonance seen by ATLAS and CMS and show that the $S\to Zh$ decay rate could naturally be of similar magnitude or larger than the diphoton rate.
\end{abstract}

\maketitle

\section{Introduction}

The preliminary evidence for a new resonance $S$ with mass around 750\,GeV seen in the diphoton invariant-mass spectrum in the $\sqrt{s}=13$\,TeV LHC run by ATLAS and CMS \cite{Aaboud:2016tru,Khachatryan:2016hje} has raised the hopes for discovering a new sector of particle physics. If confirmed with higher statistics, this would allow for a direct exploration of physics beyond the Standard Model (SM) and could provide answers to the persistent questions about the stability of the electroweak scale and the origin of dark matter. The absence of signals for decay modes other than $S\to\gamma\gamma$ already provides important information about the properties of the new particle. In particular, the production and decay through SM particles can be excluded under reasonable assumptions, because the corresponding tree-level decay would completely dominate over the diphoton mode. Most theoretical speculations assume that the new resonance is a singlet under the SM gauge group and carries spin 0 or 2. The spin-2 option is interesting, as it might hint at a connection between gravity and the weak scale, which is provided e.g.\ by models featuring a warped extra dimension \cite{Randall:1999ee}. However, in the context of such models one would generically expect the existence of lighter states with lower spin, unless the curvature of the extra dimension is of trans-Planckian size \cite{Carmona:2016jhr}. 

Let us suppose, then, that the new resonance $S$ is a gauge-singlet, spin-0 particle. Since its mass is much larger than the electroweak scale, its interactions can be described in terms of local operators in the unbroken phase of the electroweak gauge symmetry. At the renormalizable level, the only interactions of $S$ with SM particles can arise from the Higgs portals
\begin{equation}
   {\cal L}_{\rm portal} 
   = - \lambda_1\,S\,\phi^\dagger\phi - \frac{\lambda_2}{2}\,S^2\,\phi^\dagger\phi \,,
\end{equation}
where $\phi$ is the Higgs doublet. The coefficient $\lambda_1$ is strongly constrained by the existing bounds on the two-particle decay modes $S\to ZZ$, $S\to WW$ and $S\to t\bar t$, which can proceed at tree level via the mass mixing of $S$ and $h$ induced by this operator, and by the bound on $S\to hh$ \cite{Bauer:2016lbe,Dawson:2016ugw}. The portal coupling $\lambda_2$, on the other hand, does not give rise to dangerous effects; its phenomenology has been explored in \cite{Carmona:2016qgo}. It is therefore a challenge to model building to find ways of suppressing the coupling $\lambda_1$, either by means of a symmetry or dynamically. In particular, a discrete $Z_2$ symmetry under which $S$ changes sign would enforce $\lambda_1=0$. If the ultraviolet theory is (at least approximately) CP invariant, then neutral particles can be classified as CP eigenstates. If $S$ is a CP-odd pseudoscalar ($J^{PC}=0^{-+}$), then $\lambda_1$ must be zero. A nice example of a dynamical suppression is provided by models in which $S$ is identified with the lowest mode of a $Z_2$-odd bulk scalar field in a warped extra dimension \cite{Bauer:2016lbe,Csaki:2016kqr}. When the Higgs sector is localized on the infrared brane, its coupling to $S$ is either suppressed by a small wave-function overlap or by a loop factor. Here we entertain the first possibility of eliminating the portal coupling $\lambda_1$ by supposing that $S$ is a CP-odd pseudoscalar, e.g.\ an axion-like particle \cite{Pilaftsis:2015ycr,Higaki:2015jag,Ben-Dayan:2016gxw,Barrie:2016ntq,Chiang:2016eav,Gherghetta:2016fhp,Dimopoulos:2016lvn}.

Probing the spin and the CP properties of the new resonance will be of high priority, if the diphoton excess is confirmed by further data. Measurements of angular distributions in $S\to ZZ\to 4l$ or $S\to Z\gamma\to 4l$ decays have been considered \cite{Chala:2016mdz,Franceschini:2016gxv}, in complete analogy to the corresponding measurements in Higgs decays \cite{Soni:1993jc}. However, in contrast to the case of the Higgs boson, the rates for these decays are likely to be much smaller than the diphoton decay rate, and hence it may require very large statistics to perform these analyses. 

In this Letter we propose the decay $S\to Zh$, which is strictly forbidden for a CP-even scalar, as a novel and independent way to test the spin and CP quantum numbers of the new particle. The very existence of this decay would constitute a smoking-gun signal for a pseudoscalar nature of $S$ (or for significant pseudoscalar couplings, in case $S$ is a state with mixed CP quantum numbers), without the need to analyze angular distributions. The observation of this decay would also exclude a spin-2 explanation of the diphoton excess \cite{Kim:2015vba}. To the best of our knowledge, this signature has been overlooked in the literature so far. Established experimental analyses searching for pseudoscalar particles decaying into $Zh$ in the context of two-Higgs-doublet models can be adapted for the proposed search.

\section{Effective Lagrangian Analysis}
\label{sec:pheno}

At the level of dimension-5 operators, the most general couplings of a CP-odd scalar to gauge bosons read
\begin{equation}
   {\cal L}_{\rm eff}^{\rm gauge}
   = \frac{\tilde c_{gg}}{M}\,\frac{\alpha_s}{4\pi}\,S\,G_{\mu\nu}^a \widetilde G^{\mu\nu,a}
    + \dots \,,
\end{equation}
where $M$ denotes the new-physics scale, and the dots represent analogous couplings to the $SU(2)_L$ and $U(1)_Y$ gauge bosons. Via this operator the resonance $S$ can be produced in gluon fusion at the LHC. The most general dimension-5 couplings of $S$ to fermions have the same form as the SM Yukawa interactions times $S/M$, and with the Yukawa matrices replaced by some new matrices. In any realistic model these couplings must have a hierarchical structure in the mass basis in order to be consistent with the strong constraints from flavor physics \cite{Goertz:2015nkp}. It is thus reasonable to assume that the dominant couplings are those to the third-generation quarks, which for a pseudoscalar $S$ and in unitary gauge can be parameterized in the form
\begin{equation}\label{cttdef}
   {\cal L}_{\rm eff}^{\rm ferm} 
   \ni - \tilde c_{tt}\,\frac{m_t}{M} \left( 1 + \frac{h}{v} \right) S\,\bar t\,i\gamma_5\,t 
    + [t\to b] \,.
\end{equation}
Via the second term the resonance $S$ can be produced in bottom-quark fusion at the LHC \cite{Franceschini:2015kwy,Gao:2015igz}.

When using an effective Lagrangian to describe the production and decays of the resonance $S$ one should keep in mind that, in many new-physics scenarios addressing the diphoton anomaly, the masses of the heavy particles which are integrated out are in the TeV range. Otherwise it is difficult to account for the relatively large diphoton signal $\sigma(pp\to S\to\gamma\gamma)=(4.6\pm1.2)$\,fb \cite{Buttazzo:2015txu}. When there is no significant mass gap between $S$ and the new sector containing these particles, then contributions from operators with dimension $D\ge 6$ are not expected to be strongly suppressed compared with those shown above. Some of these operators induce new structures not present at dimension-5 level. 

\subsection{\boldmath Operator analysis of $S\to Zh$ decay}
\label{sec:SZh}

\begin{figure}
\includegraphics[width=0.47\textwidth]{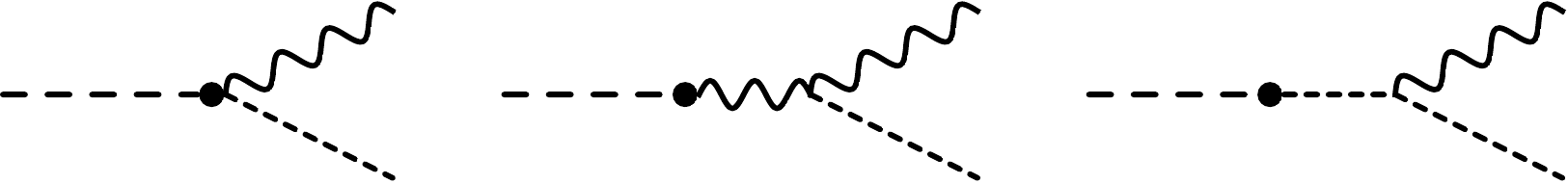}
\caption{\label{fig:EoMgraphs} 
Tree-level diagrams representing the contribution of the operator in (\ref{O5cand}) to $S\to Zh$ decay. The internal dashed line in the third graph represents the Goldstone boson $\varphi_3$.}
\end{figure}  
  
The decay $S\to Zh$ has been studied in the context of two-Higgs-doublet models, where it arises at renormalizable level via the kinetic terms \cite{Baer:1992uu,Kominis:1994fa}. However, this requires the pseudoscalar $S$ to be light (since the effect vanishes in the decoupling limit) and carry electroweak quantum numbers. For the case of a gauge-singlet scalar considered here, the effective Lagrangian up to dimension~5 does not contain any polynomial operator which could mediate the decay $S\to Zh$ at tree level. The obvious candidate
\begin{equation}\label{O5cand}
   (\partial^\mu S) \left( \phi^\dagger iD_\mu\,\phi + \mbox{h.c.} \right) 
   \to - \frac{g}{2c_w}\,(\partial^\mu S)\,Z_\mu\,(v+h)^2 \,,
\end{equation}
where $c_w\equiv\cos\theta_w$ and the last expression holds in unitary gauge, can be reduced to operators containing fermionic currents using the equations of motion. This is a consequence of the partial conservation of the Higgs current, 
\begin{equation}
   \vspace{-1.5mm}
   \partial^\mu\!\left( \phi^\dagger iD_\mu\,\phi + \mbox{h.c.} \right)
   \to - \Big( 1 + \frac{h}{v} \Big) \sum_f\,2T_3^f m_f \bar f\,i\gamma_5 f \,,
\end{equation}
where $T_3^f$ is the third component of weak isospin. The resulting operators are of the same form as those in (\ref{cttdef}) and do not give rise to a tree-level $S\to Zh$ matrix element. Indeed, adding up the diagrams shown in Figure~\ref{fig:EoMgraphs} one finds that the tree-level $S\to Zh$ matrix element of the operator in (\ref{O5cand}) vanishes identically, and the same is true for the $S\to Zhh$ matrix element.
\footnote{In \cite{Gonzalez-Alonso:2014rla} the operator in (\ref{O5cand}) was used to illustrate new-physics effects which could induce the Higgs decay $h\to Z\phi$ into a hypothetical, light scalar particle $\phi$. However, we find that its contribution vanishes when all graphs shown in Figure~\ref{fig:EoMgraphs} are included.} 
Importantly, however, in extensions of the SM containing heavy particles whose masses arise (or receive their dominant contributions) from electroweak symmetry breaking, the non-polynomial operator
\begin{equation}\label{Obeauty}
   O_5 = (\partial^\mu S) \left( \phi^\dagger iD_\mu\,\phi + \mbox{h.c.} \right)
    \ln\frac{\phi^\dagger\phi}{\mu^2} 
\end{equation}
can be induced \cite{Pierce:2006dh}. Using an integration by parts and the equations of motion, and neglecting fermionic terms which do not contribute to $S\to Zh$ decay at tree level, this operator can be reduced to 
\begin{equation}\label{beauty}
\begin{aligned}
   O_5 &\,\hat =\, - S \left( \phi^\dagger iD_\mu\,\phi + \mbox{h.c.} \right)
    \frac{\partial^\mu (\phi^\dagger\phi)}{\phi^\dagger\phi} \\
   &\!\to \frac{g}{c_w}\,S\,Z_\mu\,(v+h)\,\partial^\mu h \,.
\end{aligned}
\end{equation}
This gives rise to non-vanishing $S\to Zh$ and $S\to Zhh$ matrix elements. At one-loop order, the $S\to Zh$ decay amplitude also receives a contribution from an operator containing quark fields, and since the Higgs boson couples proportional to the quark mass it suffices to consider the term involving the top quark. The complete dimension-5 Lagrangian is therefore
\begin{equation}\label{Leff5}
   {\cal L}_{\rm eff}^{D=5} 
   = \frac{C_5}{M}\,O_5 + \frac{c_5^t}{M}\,S \left( i\bar Q_L\tilde\phi\,t_R + \mbox{h.c.} \right) ,
\end{equation}
where $Q_L$ is the third-generation left-handed quark doublet and $\tilde\phi=i\sigma_2\phi^*$. Comparison with (\ref{cttdef}) shows that the coefficient $c_5^t$ is given by $c_5^t=-y_t\,\tilde c_{tt}$.

The operator $O_5$ is absent in models where the new heavy particles have masses not related to the  electroweak scale. Also, as we will show below, the one-loop matrix element of the operator multiplying the Wilson coefficient $c_5^t$ is suppressed by a factor $m_t^2/m_S^2$. It is therefore worthwhile to include operators of higher dimension in the effective Lagrangian. At dimension~7 there is a single operator giving rise to a tree-level contribution to the $S\to Zh$ amplitude. It reads
\begin{equation}
\begin{aligned}
   O_7 &= (\partial^\mu S) \left( \phi^\dagger iD_\mu\,\phi + \mbox{h.c.} \right)
    \phi^\dagger\phi \\
   &\hspace{1mm}\hat{=}\, - S \left( \phi^\dagger iD_\mu\,\phi + \mbox{h.c.} \right)
    \partial^\mu (\phi^\dagger\phi) \,,
\end{aligned}
\end{equation}
which differs from the operator in (\ref{beauty}) by a factor $\phi^\dagger\phi$. In the second line we have again neglected operators containing fermions. At one-loop order there exist several operators contributing to the decay $S\to Zh$. Those relevant to our analysis are
\begin{equation}\label{Leff7}
\begin{aligned}
   {\cal L}_{\rm eff}^{D\le 7} 
   &= \frac{C_7}{M^3}\,O_7 + \frac{c_6^t}{M^2}\,
    \bar t_R\,\tilde\phi^\dagger i\rlap{\,/}{D}\,\tilde\phi\,t_R \\
   &\quad\mbox{}+ \frac{c_{7a}^t}{M^3}
    \left[ i S\,\bar Q_L i\rlap{\,/}{D}\,i\rlap{\,/}{D}\,\tilde\phi\,t_R + \mbox{h.c.} \right] \\
   &\quad\mbox{}+ \frac{c_{7b}^t}{M^3}\,(\partial^\mu S)\,
    \bar t_R\,\tilde\phi^\dagger\gamma^\mu\tilde\phi\,t_R + \dots \,,
\end{aligned}
\end{equation}
plus analogous operators containing the right-handed bottom quark. The dimension-6 operator proportional to $c_6^t$ contributes in conjunction with the operator multiplying $c_5^t$ in (\ref{Leff5}) to give a contribution of order $c_5^t c_6^t/M^3$. Below we will consider a concrete new-physics model containing heavy vector-like fermions, in which the operators shown above arise in the low-energy effective Lagrangian. However, we find it instructive to focus first on the contributions from the tree-level matrix elements of the operators $O_5$ and $O_7$, and on the dimension-5 contribution induced by top-quark loops.

\subsection{Potential tree-level contributions}

The tree-level matrix elements of the operators $O_5$ and $O_7$ give rise to the decay amplitude \begin{equation}
   i{\cal A}(S\to Zh) = - \frac{2m_Z\,\epsilon_Z^*\cdot p_h}{M}
    \left( C_5 + \frac{v^2}{2M^2}\,C_7 \right) .
\end{equation}
The $Z$ boson is longitudinally polarized, and hence the structure $2m_Z\,\epsilon_Z^*\cdot p_h\approx 2p_Z\cdot p_h\approx m_S^2$ is proportional to the mass squared of the heavy singlet. The decay rate is given by
\begin{equation}
   \Gamma(S\to Zh) = \frac{m_S^3}{16\pi M^2}\,\Big| C_5 + \frac{v^2}{2M^2}\,C_7 \Big|^2
   \lambda^{3/2}(1,x_h,x_Z) \,,
\end{equation}
where we have defined $x_i=m_i^2/m_S^2$ and $\lambda(x,y,z)=(x-y-z)^2-4yz$. With an ${\cal O}(1)$ value of the coefficient $C_5$ and a new-physics scale $M=1$\,TeV, this partial width is of order 7\,GeV. If $C_5$ vanishes and $C_7={\cal O}(1)$, the width is of order 7\,MeV. The current experimental upper bound on the $pp\to S\to Zh$ cross section times branching ratio at $\sqrt{s}=13$\,TeV is 73\,fb \cite{ATLAS-CONF-2016-015}, which is more than an order of magnitude larger than the observed $S\to\gamma\gamma$ rate. If we assume for simplicity that the resonance is produced in gluon fusion, and that its dominant decay is into dijets ($S\to gg$), this upper bound translates to \cite{Bauer:2016lbe}
\begin{equation}\label{C5bound}
   \Big| C_5 + \frac{v^2}{2M^2}\,C_7 \Big| < 0.031\,\frac{M}{\text{TeV}} \,.
\end{equation}
This is a rather strong bound, which suggests that the $C_5$ coefficient (if present) should be loop suppressed in realistic models. If $C_5$ vanishes, then the bound translates to $|C_7|<1.02\,(M/{\rm TeV})^3$. Even in this latter case it is conceivable that the $S\to Zh$ decay mode has a significantly larger rate than the diphoton mode, in which case it should be possible to observe it in the near future.

\subsection{\boldmath $D=5$ contribution induced by fermion loops}

\begin{figure}
\includegraphics[width=0.47\textwidth]{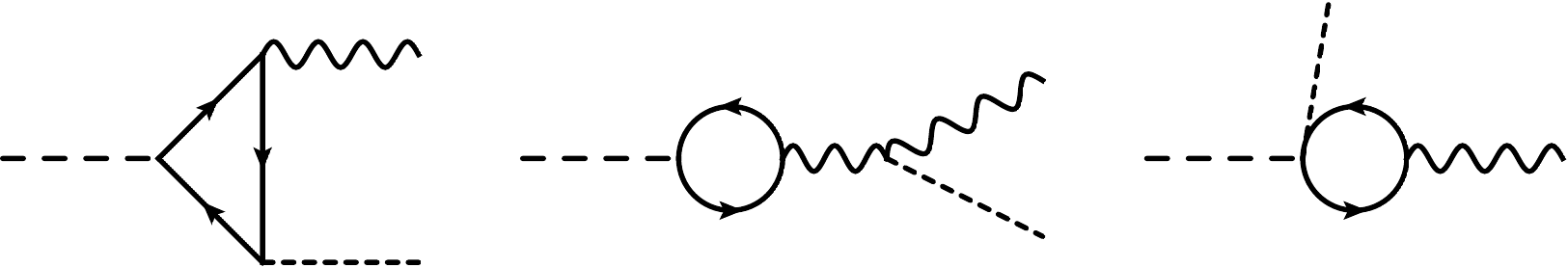}
\caption{\label{fig:toploops} 
Top-loop contributions to $S\to Zh$ decay. We omit a mirror copy of the first graph with a different orientation of the fermion loop and diagrams involving Goldstone bosons.}
\end{figure}

The leading dimension-5 contribution induced by top-quark loops in the low-energy effective theory arises from the second operator in (\ref{Leff5}), with coefficient $c_5^t=-y_t\,\tilde c_{tt}$. The corresponding Feynman diagrams are shown in Figure~\ref{fig:toploops}. We have evaluated these diagrams in a general $R_\xi$ gauge. The resulting decay amplitude is
\begin{equation}\label{ourresult}
   i{\cal A}(S\to Zh) = \frac{2m_Z\,\epsilon_Z^*\cdot p_h}{M}\,
    \frac{N_c\,y_t^2}{8\pi^2}\,T_3^t\,\tilde c_{tt}\,F \,,
\end{equation}
where $T_3^t=\frac12$, and $F$ denotes the parameter integral
\begin{equation}\label{Fres}
   F = \int_0^1\!d[xyz]\,\frac{2m_t^2-x m_h^2-z m_Z^2}{m_t^2-xz m_S^2-xy m_h^2-yz m_Z^2-i0} \,,
\end{equation}
with $d[xyz]\equiv dx\,dy\,dz\,\delta(1-x-y-z)$. The factor $y_t^2$ in (\ref{ourresult}) ensures that analogous contributions from light fermions in the loop are negligible. Evaluating the integral for $m_t\equiv m_t(m_S)=146.77$\,GeV and with the physical Higgs and $Z$-boson masses gives $F\approx-0.009+ 0.673\,i$. 
It is instructive to study the behavior of the function $F$ in more detail, neglecting for simplicity the small effects due to $m_h^2$ and $m_Z^2$. In the limit $m_t^2\ll m_S^2$, we obtain
\begin{equation}
   F = - \frac{m_t^2}{m_S^2} \left( \ln\frac{m_S^2}{m_t^2} - i\pi \right)^2
    + {\cal O}\bigg( \frac{m_t^4}{m_S^4} \bigg) \,.
\end{equation}
This function is formally suppressed by a factor $m_t^2/m_S^2$, and its real part is accidentally small. The imaginary part, on the other hand, is enlarged by a factor $2\pi\ln(m_S^2/m_t^2)$, and as a result $|F|$ is numerically of ${\cal O}(1)$. 
%
%
If the dominant contribution to the $S\to Zh$ decay amplitude is indeed related to the top-quark contribution proportional to $c_5^t$, then we can derive a relation between the $S\to Zh$ and $S\to t\bar t$ rates. It reads
\begin{equation}\label{eq17}
   \frac{\Gamma(S\to Zh)}{\Gamma(S\to t\bar t)}
   = \frac{3y_t^2}{16\pi^2} \left( \frac{m_S}{4\pi v} \right)^2 \left| F \right|^2
    \frac{\lambda^{3/2}(1,x_h,x_Z)}{\sqrt{1-4x_t}} \,,
\end{equation}
which evaluates to $3.6\cdot 10^{-4}$. The present experimental upper bound on the $S\to t\bar t$ rate of about 0.7\,pb at $\sqrt{s}=8$\,TeV \cite{Aad:2015fna} yields $\sigma(pp\to S\to t\bar t)<3.2$\,pb at 13\,TeV under the assumption of gluon-initiated production. Relation (\ref{eq17}) then implies $\sigma(pp\to S\to Zh)_{\rm top}<1.2$\,fb. 

It is interesting to consider the hypothetical limit where one takes the fermion mass $m_t$ in (\ref{Fres}) much larger than the mass of the resonance $S$, i.e.\ $m_t^2\gg m_S^2$. In this case the parameter integral yields $F=1+{\cal O}(m_S^2/m_t^2)$. The fermion is a very heavy particle, which should be integrated out from the low-energy theory. The contribution (\ref{ourresult}) then corresponds to a matching contribution to the Wilson coefficient of a local dimension-5 operator, suppressed by only a single power of $M$. Close inspection shows that the leading term corresponds to a matching contribution to the operator $O_5$ in (\ref{Obeauty}). The non-polynomial structure arises because the particle integrated out (the hypothetical heavy fermion) receives its mass from electroweak symmetry breaking, so it is heavy only in the broken phase of the theory. The equivalent form of the operator shown in (\ref{beauty}) can readily be mapped onto the structure of the parameter integral in (\ref{Fres}). Consider, as an illustration, a sequential fourth generation of heavy leptons, and assume that the heavy charged state $L$ has mass $m_L>m_S/2$ and a coupling $\tilde c_{LL}$ to the pseudoscalar resonance defined in analogy to (\ref{cttdef}). Integrating out this heavy lepton generates the contribution
\begin{equation}
   C_5 = \frac{y_L^2\,\tilde c_{LL}}{16\pi^2} 
   = \frac{m_L^2\,\tilde c_{LL}}{8\pi^2 v^2} > 0.03\,\tilde c_{LL}
\end{equation}
to the Wilson coefficient of the operator $O_5$. Comparison with (\ref{C5bound}) indicates that, for $\tilde c_{LL}={\cal O}(1)$ of natural size, it would be possible in this case to obtain a $S\to Zh$ decay rate close to the present experimental upper bound. 

There is an interesting subtlety related to the calculation of $F$ worth pointing out. We have obtained the result (\ref{ourresult}) using the naive definition of $\gamma_5$, such that $\{\gamma^\mu,\gamma_5\}=0$. It is well know that this scheme is not consistent beyond tree level. We have thus repeated the calculation using the `t\,Hooft-Veltman (HV) scheme \cite{'tHooft:1972fi}, in which $\gamma_5$ anticommutes with $\gamma^\mu$ for $\mu=0,1,2,3$, while it commutes with the remaining $(d-4)$ $\gamma^\mu$ matrices. We then find an additional, gauge-dependent contribution to $F$ given by
\begin{equation}\label{delFHV}
   \delta F_{\rm HV} = - 1 - \frac23\,\frac{6m_t^2-m_S^2}{m_S^2-\xi m_Z^2} \,.
\end{equation}
Note the peculiar feature that in unitary gauge ($\xi=\infty$) this contribution would have the effect of subtracting the leading asymptotic contribution to $F$ in the limit $m_t^2\gg m_S^2$, leaving a result which formally corresponds to the matrix element of a dimension-7 operator. This seems to contradict the conclusion drawn above. However, it is well known that the HV scheme (like any other consistent scheme for implementing $\gamma_5$ in dimensional regularization) violates the chiral Ward identities of the electroweak theory \cite{Bonneau:1980ya}. In our case, the relevant Ward identity takes the form
\begin{equation}
   k_\mu \Gamma^\mu(k) = -im_Z\,\Gamma(k) \,,
\end{equation}
where $\Gamma^\mu(k)$ is the proper vertex function of an on-shell $S$ decaying to an on-shell Higgs boson and a $Z$-boson current with momentum $k$, while $\Gamma(k)$ is the corresponding proper vertex function with the current replaced by the Goldstone boson $\varphi_3$. The Ward identity must be restored by means of appropriate counterterms. We find that, when this is done consistently, the counterterm contribution to the $S\to Zh$ decay amplitude precisely cancels the extra term in (\ref{delFHV}), so that we recover the result obtained using the naive definition of $\gamma_5$. This finding should not come as a surprise. In \cite{Korner:1991sx} a consistent scheme for implementing $\gamma_5$ in dimensional regularization was proposed, which for traces involving an even number of $\gamma_5$ matrices yields results identical to those obtained in the naive scheme with anticommuting $\gamma_5$.

\section{Heavy vector-like fermions}

It is instructive to illustrate our findings with a concrete new-physics model, which generates the effective interactions of the scalar resonance with SM particles via loop diagrams involving heavy vector-like fermions that are mixed with the SM fermions. Such a scenario is realized, e.g., in models of partial compositeness or warped extra dimension \cite{Kaplan:1991dc,Grossman:1999ra,Gherghetta:2000qt}. We consider an $SU(2)_L$ doublet $\psi=(T~B)^T$ of vector-like quarks with hypercharge $Y_\psi=\frac16$, which mixes with the third-generation quark doublet of the SM. The most general Lagrangian is
\begin{align}
   {\cal L} &= \bar\psi\left( i\rlap{\,/}{D} - M \right) \psi + \bar Q_L\,i\rlap{\,/}{D}\,Q_L
    + \bar t_R\,i\rlap{\,/}{D}\,t_R + \bar b_R\,i\rlap{\,/}{D}\,b_R \nonumber\\
   &\quad\mbox{}- y_t \big( \bar Q_L\tilde\phi\,t_R + \mbox{h.c.} \big)
    - \big( g_t \bar\psi\,\tilde\phi\,t_R + g_b \bar\psi\,\phi\,b_R + \mbox{h.c.} \big) \nonumber\\
   &\quad\mbox{}- c_1 S\,\bar\psi\,i\gamma_5\,\psi 
    - i c_2 S \big( \bar Q_L\psi - \bar\psi\,Q_L \big) \,,
\end{align}
where we neglect the small Yukawa coupling $|y_b|\ll 1$ of the bottom quark. The terms in the last line contain the couplings to the pseudoscalar resonance $S$. The mass mixing induced by the couplings $g_i$ leads to modifications of the masses and Yukawa couplings of the SM top and bottom quarks by small amounts of order $g_i^2\,v^2/M^2$. Likewise, the masses of the heavy $T$ and $B$ quarks are split by a small amount $M_T-M_B\approx (g_t^2-g_b^2)\,v^2/(4M)$. 

Integrating out the heavy fermion doublet at tree level by solving its equations of motion, we generate the operators in the effective Lagrangians (\ref{Leff5}) and (\ref{Leff7}) with coefficients $c_5^t=c_2\,g_t=-y_t\,\tilde c_{tt}$ and 
\begin{equation}\label{c6c7res}
   c_6^f = g_f^2 \,, \qquad c_{7a}^f = c_2\,g_f \,, \qquad c_{7b}^f = c_1\,g_f^2 \,,
\end{equation}
for $f=t,b$. The coefficient $c_6^b$ is constrained from precision measurements of the $Z$-boson couplings to fermions performed at LEP and SLD. A recent global analysis of electroweak precision tests finds \cite{Efrati:2015eaa}
\begin{equation}
   c_6^b = g_b^2 = (0.76\pm 0.27) \left( \frac{M}{\rm TeV} \right)^2 ,
\end{equation}
where the pull away from zero is largely driven by the $b$-quark forward-backward asymmetry $A_b^{\rm FB}$, whose experimental value is about 2.8$\sigma$ smaller than the SM prediction \cite{ALEPH:2005ab}. Our model can resolve this anomaly in a natural way. It is likely that the coupling $g_t$ is at least as large as $g_b$, perhaps even significantly larger. In fact, in our model the relation $\tilde c_{bb}/\tilde c_{tt}=(g_b/g_t)\,(m_t/m_b)$ holds, and hence the coupling of the resonance $S$ to bottom quarks defined in (\ref{cttdef}) can be rather large.

The coefficient $C_7$ in (\ref{Leff7}) is induced at one-loop order by diagrams such as those shown in Figure~\ref{fig:toploops}, where now both heavy and light quarks can propagate in the loops. In order to calculate $C_7$ a proper matching onto the low-energy theory must be performed. We obtain
\begin{widetext}
\begin{equation}\label{C7result}
\begin{aligned}
   \frac{v^2}{2}\,C_7 &= c_1 \sum_{f=t,b}\,\frac{N_c\,g_f^2}{16\pi^2}\,\bigg\{
    2 T_3^f\,\bigg[ m_f^2 \left( L - \frac32 \right) - \frac{m_h^2}{12} + \frac{m_Z^2}{36} 
    + \frac{g_f^2\,v^2}{4} \bigg] - \frac23\,Q_f s_w^2 m_Z^2 \left( L - \frac32 \right)\! \bigg\} \\
   &\quad\mbox{}+ \tilde c_{tt}\,\frac{N_c\,y_t^2}{16\pi^2}\,\bigg\{ 
    2 T_3^t \left[ 3m_t^2 \left( L - \frac32 \right) - \frac{m_h^2}{2}\,\Big( L-\frac76 \Big) 
    - \frac{m_Z^2}{6} \left( L + \frac{19}{6} \right) - g_t^2 v^2 \left( L - \frac94 \right) \right] 
    + Q_t s_w^2 m_Z^2 \bigg\} \,,
\end{aligned}
\end{equation}
\end{widetext}
where $L=\ln(M^2/\mu^2)$. There is a non-trivial operator mixing at dimension-7 order, such that the scale dependence of the coefficient $C_7$ cancels against the scale dependence of one-loop matrix elements in the low-energy effective theory. 
We now demonstrate this cancellation for the terms proportional to $c_1$ shown in the first line. It follows from (\ref{c6c7res}) that for this purpose we need to calculate the matrix element of the operator multiplying $c_{7b}^t$ and $c_{7b}^b$ in (\ref{Leff7}). At one-loop order we find the fermion-loop contribution
\begin{equation}\label{myresult}
\begin{aligned}
   &i{\cal A}_{\rm ferm} = - 2 m_Z\,\epsilon_Z^*\cdot p_h\,\frac{c_1}{M^3}\,\frac{N_c}{16\pi^2} \\
   &\times \bigg\{ g_t^2\,\bigg[ 2T_3^t\,m_t^2 \left( \ln\frac{\mu^2}{m_t^2} + 2 - 2 f_Z - f_h + f_S 
    + \frac{F'}{2} \right) \\
   &\hspace{6mm}\mbox{}- \frac23\,Q_t s_w^2 m_Z^2 \left( \ln\frac{\mu^2}{m_t^2} 
    + \frac53 - 2 f_Z + \frac{4m_t^2}{m_Z^2} \left( 1 - f_Z \right) \right) \!\bigg] \\
   &\hspace{6mm}\mbox{}- \frac23\,g_b^2\,Q_b s_w^2 m_Z^2 \left( \ln\frac{\mu^2}{m_Z^2} 
    + \frac53 + i\pi \right)\! \bigg\} 
    + {\cal O}\bigg(\frac{\tilde c_{tt}\,y_t^2}{16\pi^2}\bigg) ,
\end{aligned}
\end{equation}
where
\begin{align}
   f_i &= \sqrt{\tau_i-1}\,\arcsin\frac{1}{\sqrt{\tau_i}} \,; \quad
    \tau_i = \frac{4m_t^2}{m_i^2} - i0 \,, \\
   F' &= \int_0^1\!d[xyz]\,
    \frac{x(m_S^2+m_h^2)+(1-x)m_Z^2-4m_t^2}{m_t^2-xz m_S^2-xy m_h^2-yz m_Z^2-i0} \,. \nonumber
\end{align}
The expression for $f_S$ must be obtained by analytic continuation, since $\tau_S<1$. Note that the $\mu$-dependent terms in (\ref{myresult}) precisely cancel the scale dependence of $C_7$ in the combination 
\begin{equation}
   i{\cal A}(S\to Zh) \big|_{\rm D=7} 
   =  - \frac{2m_Z\,\epsilon_Z^*\cdot p_h}{M^3}\,\frac{v^2}{2}\,C_7 + i{\cal A}_{\rm ferm} \,.
\end{equation}
To estimate the dimension-7 contribution we set $\mu=m_Z$ in (\ref{C7result}) and neglect the fermion-loop contributions in the low-energy theory. All large logarithms $L\approx 4.2$ are included in the Wilson coefficient $C_7$, for which we obtain
\begin{equation}
\begin{aligned}
   C_7 &\approx \Big[ c_1 \left( 3.48\,g_t^2 + 0.95\,g_t^4 + 0.14\,g_b^2 - 0.95\,g_b^4 \right) \\
   &\hspace{5.5mm}\mbox{}+ \tilde c_{tt} \left( 6.44 - 5.34\,g_t^2 \right) \!\Big]\cdot 10^{-2} \,.
\end{aligned}
\end{equation}
For natural values of the couplings this coefficient can be of order a few to tens of percent. For example, with $g_t=g_b=1$ we have $C_7\approx(0.036\,c_1+0.011\,\tilde c_{tt})$, while for $g_t=2$ and $g_b=0.5$ we get $C_7\approx(0.29 \,c_1-0.15\,\tilde c_{tt})$. If the resonance $S$ is produced in gluon fusion and predominantly decays into dijets, one obtains a $pp\to S\to Zh$ rate of 1\,fb or 10\,fb for $|C_7|=(0.12$ or $0.38) \times (M/{\rm TeV})^3$, see (\ref{C5bound}). If the mass of the vector-like quarks is $M\lesssim 1$\,TeV, such rates can naturally be obtained in our model.

\section{Conclusions}

We have presented the first detailed analysis of the decay $S\to Zh$ of a gauge-singlet, heavy scalar resonance $S$ and pointed out that this decay is allowed only if $S$ has CP-odd couplings. Such a scalar boson arises in a variety of models in which the SM is connected to a new, hidden sector via Higgs-portal interactions. The decay $S\to Zh$ can then be used to determine the CP nature of the new state. Using a model-independent analysis based on an effective Lagrangian, we have shown that the decay amplitude receives contributions starting at dimension~5. These come either from top-quark loop diagrams or from a non-polynomial local operator $O_5$, which can only be induced upon integrating out heavy particles whose masses arise from electroweak symmetry breaking. If such particles do not exist, then a tree-level contribution can first arise at dimension-7 order and can be parameterized in terms of a unique operator $O_7$. 

While our analysis is completely general, we have illustrated our results with the example of the 750\,GeV diphoton resonance seen by ATLAS and CMS, for which we have discussed new-physics scenarios that can give rise to a production times decay rate exceeding that for the diphoton decay. In a model featuring a sequential fourth generation of heavy leptons, the $S\to Zh$ rate can be close to its present experimental bound of 73\,fb. Perhaps more interestingly, in a weakly coupled model containing a heavy vector-like fermion doublet transforming like the left-handed quark doublet of the SM, we can at the same time explain the persistent anomaly of the $b$-quark forward-backward asymmetry on the $Z$ pole and obtain a $S\to Zh$ rate in the $1\!-\!10$\,fb range. If the diphoton resonance is confirmed by future data, it should be possible to see the $S\to Zh$ decay mode if $S$ is a CP-odd particle.

\acknowledgments
We are grateful to M.~Beneke, J.~Henn, J.~Thaler and A.~von Manteuffel for useful discussions. M.B.\ acknowledges the support of the Alexander von Humboldt Foundation. The work of M.N.\ and A.T.\ is supported by the Advanced Grant EFT4LHC of the European Research Council (ERC), the DFG Cluster of Excellence PRISMA (EXC 1098) and grant 05H12UME of the German Federal Ministry for Education and Research (BMBF). We are grateful to KITP Santa Barbara, MITP Mainz, Universit\`a di Napoli Federico II and INFN for hospitality and support during different stages of this work.
\vfil

\end{document}